# Assessing clinical utility of Machine Learning and Artificial Intelligence approaches to analyze speech recordings in Multiple Sclerosis: A Pilot Study

Svoboda, E.[1,2], Bořil, T., PhD[1], Rusz, J., [3,4] Tykalová, T.[3], Horáková, D. [4], Guttmann, C.R.G., MD[9], Blagoev, K.B., PhD[8], Hatabu, H., MD, PhD[6], Valtchinov, V.I., PhD[5,6,7]

[1]Institute of Formal and Applied Linguistics, Faculty of Mathematics and Physics, Charles University, Prague, Czech Republic
[2]Institute of Phonetics, Faculty of Arts, Charles University, Prague, Czech Republic
[3]Department of Circuit Theory, Faculty of Electrical Engineering, Czech Technical University in Prague, Prague, Czech Republic
[4]Department of Neurology and Center of Clinical Neuroscience, First Faculty of Medicine, Charles University and General University Hospital, Prague, Czech Republic

[5]Center for Evidence-Based Imaging, [6]Department of Radiology, Brigham and Women's Hospital, Harvard Medical School, Boston, MA, USA
[7]Department of Biomedical Informatics, Harvard Medical School, Boston, MA, USA
[8]Department of Biophysics, Johns Hopkins University, Baltimore, MD 21218, USA
[9]Center for Neurological Imaging, Brigham & Women's Hospital
and Harvard Medical School

**Corresponding Author:**

Vladimir Valtchinov, PhD
Center for Evidence Based Imaging (CEBI)
20 Kent Street
Brookline, MA 02445
Phone: +1.617.525.9711
Fax: +1.617.525.7575
e-mail: vvaltchinov@bwh.harvard.edu

**Abstract**

Background: An early diagnosis together with an accurate disease progression monitoring of multiple sclerosis is an important component of successful disease management. Prior studies have established that multiple sclerosis is correlated with speech discrepancies. Early research using objective acoustic measurements has discovered measurable dysarthria.

Objective: To determine the potential clinical utility of machine learning and deep learning/AI approaches for the aiding of diagnosis, biomarker extraction and progression monitoring of multiple sclerosis using speech recordings.

Methods: A corpus of 65 MS-positive and 66 healthy individuals reading the same text aloud was used for targeted acoustic feature extraction utilizing automatic phoneme segmentation. A series of binary classification models was trained, tuned, and evaluated regarding their Accuracy and area-under-curve.

Results: The Random Forest model performed best, achieving an Accuracy of 0.82 on the validation dataset and an area-under-curve of 0.76 across 5 k-fold cycles on the training dataset. 5 out of 7 acoustic features were statistically significant.

Conclusion: Machine learning and artificial intelligence in automatic analyses of voice recordings for aiding MS diagnosis and progression tracking seems promising. Further clinical validation of these methods and their mapping onto multiple sclerosis progression is needed, as well as a validating utility for English-speaking populations.

**Introduction**

It is almost universally accepted that MS, a chronic inflammatory autoimmune neurological disease of the CNS, is due to the changes of the CNS myelinated axons, creating inflammatory plaques that effectively cause demyelination with axonal transection. (1,2)

The clinical development and manifestation course of MS is highly varied and unpredictable. In most MS patients, episodes of reversible neurological deficits is often followed by a progressive neurological deterioration. (2) Epidemiologically, MS affects at least 14,908 people in the Czech Republic and an estimated 620,000-720,000 people in the US. (3) It typically presents in young adults (mean age of onset, 20-30 years); it is expected that up to one-half of subjects will need physical help to walk within 15 years after the onset of the disease. (2,4)

There is no single diagnostic test for MS. Diagnosis is made based on clinical evidence from multiple testing procedures, some of which are quite invasive: a combination of presentation signs and symptoms, diagnostic imaging findings (for example, MRI T2 lesions), and laboratory findings (ie CSF–specific oligoclonal bands), which are components of the 2017 McDonald Criteria. (5)

Recently, speech patterns have been shown to be a good indicator for the presence of neurological disorders, specifically in the case of MS. An early study in 1987 by Gerald et al. (6) was the first to describe the effects of this disease not only on speech, but also on linguistic capabilities in general. In a small sample of 23 individuals, they established that multiple sclerosis has noticeable effects on how afflicted individuals communicate. This study was extended by Rusz et al. in 2018, (7,8) where for the first time they introduced objective acoustic criteria showing that MS-afflicted speech differs significantly from normal speech. According to Hartelius, (9) at least some vocal impairment is perceptually present in 51% of all MS patients.

In 2021, Noffs et al. (10) demonstrated that some objective acoustic measurements of speech correlate with disability scores in MS-afflicted patients even when there is no perceivable dysarthria present, specifically intensity decay and decreased frequency variability. The sensitivity of speech disorders towards MS has not insofar been assessed using fully automated methods based on individual phoneme segmentation, nor has the potential of such a vocal fingerprint to detect and track the progression of MS been tested using machine learning methods.

To assess the utility of automated methods for speech analysis in MS patients in this study we undertook the following aims: a) create a set of acoustic parameters able to discern recordings of speakers with MS from healthy speakers; b) find out how strong are the differences between the MS-afflicted and the healthy speakers with respect to each of these parameters separately; c) test how well these parameters discriminate between these speakers using machine learning; and d) assess the feasibility of creating an automated tool for diagnosis and disease progression monitoring based on these (and potentially additional) parameters.

**Methods**

Study Setting and Human Subjects Approval

All MS patients were diagnosed with a neurologically-confirmed diagnosis of MS according to the revised McDonald Criteria. (11) All patients were relapse-free for at least 30 days prior to testing. Each patient was ranked according to the Expanded Disability Status Scale (EDSS). (12)

In addition, a healthy control group free of neurological or communication disorders was included. All participants were native speakers of the Central Bohemian dialect of Czech.

Speech recordings

Speech recordings were performed in a quiet room with a low ambient noise level using a head-mounted condenser microphone (Beyerdynamic Opus 55, Heilbronn, Germany) placed approximately 5 cm from the subject's mouth. (13) Speech signals were sampled at 48 kHz with 16-bit resolution. Each subject was recorded during a single session with a speech specialist.

All of these individuals were recorded reading out loud the same excerpt from Karel Čapek 's *Měl jsem psa a kočku* in the original Czech. (14) This text has the benefit of being cognitively and articulatorily little to moderately demanding while also utilizing the entirety of the Czech phonemic inventory. It is 230 syllables long and a healthy native speaker of Czech can be expected to read it aloud within 40 – 50 seconds.

Annotation and feature extraction

*Prague Labeller*, a *HTK*-based implementation of the Hidden Markov Model algorithm originally intended for use in phonetics, was used to automatically delimit boundaries of phoneme realizations within each of the recordings. (15) In a recording of the Czech word *Minda* /minda/, for example, this tool finds the beginning and end of the articulation of the /m/ phoneme, considering that a short silence may precede the word and thus delimiting silences as well as phonemic boundaries.

## Validation of the Automatic Algorithm for Phoneme Extraction

To validate the accuracy of *Prague Labeller*, the features extracted using the tool were correlated against features extracted from the same speakers using human experts according to rules strictly defined in *Fonetická segmentace hlásek*. (16)

W used human expert annotation that was available for the entire Control cohort and a subset of the Cases cohort, totaling 18 speakers. 7 of these features were found to be significantly correlated with expert annotation, as shown in Table 1.

In the case of one MS-speaker recording, it was discovered that *Prague Labeller* had placed the last boundary of the annotation midway through the recording, creating an artificial outlier. This recording was discarded. Pearson's r correlation coefficient was used to cross-correlate the automatically extracted features with their counterparts extracted by human experts.

## ML algorithms for Predictive Risk modelling

These vectors were combined with information about the speakers' age and gender at birth, on which a binary classification array of models was trained using the *R* programming language and the package *caret*. (17)

| Parameter | Cases | | Controls | |
| --- | --- | --- | --- | --- |
| | *p* value | corr.coeff. | *p* value | corr.co-eff. |
| **Speech duration** | **$1.9 \times 10^{-10}$** | **0.99** | **$8.9 \times 10^{-84}$** | **0.99** |
| Silence-to-speech ratio | 0.41 | 0.06 | 0.005 | 0.31 |
| **Vowel-to-speech ratio** | **$2.5 \times 10^{-5}$** | **0.78** | **$3.9 \times 10^{-26}$** | **0.91** |
| **CSI of vowel duration** | **0.01** | **0.59** | **$3.3 \times 10^{-22}$** | **0.87** |
| CSI of $f_0$ | 0.12 | 0.32 | $4.9 \times 10^{-06}$ | 0.51 |
| **Quantile difference of $f_0$** | **0.001** | **0.70** | **$2.1 \times 10^{-16}$** | **0.81** |
| **Unvoiced stop duration mean** | **$1.4 \times 10^{-7}$** | **0.93** | **$1.9 \times 10^{-39}$** | **0.97** |
| **CSI of intensity** | **$2.4 \times 10^{-27}$** | **0.99** | **$2.5 \times 10^{-84}$** | **0.99** |
| **Spectral centroid of /s/, SD** | **$5.3 \times 10^{-7}$** | **0.83** | **$1.3 \times 10^{-15}$** | **0.79** |
| Vowel F1, SD | 0.40 | 0.07 | $1.1 \times 10^{-4}$ | 0.44 |
| Vowel F2, SD | 0.41 | 0.04 | 0.42 | 0.02 |
| Vowel F3, SD | 0.53 | -0.02 | $1.1 \times 10^{-15}$ | 0.79 |

**Table 1**:
A table of Pearson correlation coefficients and associated *p* values obtained by running a correlation test on the acoustic features extracted using *Prague Labeller* and features extracted using annotation by human experts, under the hypothesis that true correlation is greater than 0. Parameters found to be significant under *p* < 0.05 for both groups are in **bold**. "$f_0$" is a shorthand for fundamental frequency.

We evaluated the performance of 7 ML algorithms in building a predictive MS risk model with acoustic features and demographic variables as independent predictors.

A common simulation and evaluation framework was set up using utilities from the *caret R* package. (18) The following algorithms were implemented and assessed: eXtreme Gradient Boosting, gbm (Generalized regression Boosting Model), GLMnet (Generalized Linear Model), KNN (k-Nearest Neighbors), Multi-Layer Perceptron Neural Network, Random Forrest (RF) and Support Vector machine with a radial kernel (SVM). We used a 5-fold cross-validation (CV) with an 80-20% balanced split for training and validation data sets. All acoustic and demographic variables were used in the model building and validation, regardless of the amount of variability in the dataset they explain, or any variable-selection procedures. The system performance was measured by the accuracy (the proportion of times the model's predictions agree with the labels of the data) using the validation set and the mean area-under-curve (AUC) for the ROC across the individual CV runs on the training set.

Univariate Statistical Analyses

In addition, to assess the statistical significance of each variable, analysis of the individual features was performed. As the dataset was found not to be normally distributed, the Kolmogorov-Smirnov two-sample test was used.

Outcome Measures

The primary outcome metrics of our analyses were the Accuracy on the holdout and the area-under-curve (AUC) for the binary classification models. Secondary outcomes were the values of the 7 acoustic and 2 demographic features extracted from the voice recording and used to construct a vector space to quantify and predict the risk of MS.

**Results**
Study Cohort

A total set of 131 matched recordings of healthy and MS individuals was recorded. A total of 65 MS patients (41 females, 24 males), with a mean age of 43.9 (standard deviation [SD] 10.5) years, mean disease duration of 14.7 (SD 8.3) years, and mean EDSS of 3.9 (SD 1.4) were recruited. Fifty-one patients were diagnosed with relapsing remitting MS, 7 with secondary progressive MS, 2 with clinically isolated syndrome and 5 with primary-progressive MS.

According to the consensus judgment of two speech-language pathologists, the dysarthria was imperceptible in 29 MS patients. The perceptible dysarthria in remaining 36 MS patients mainly featured a combination of spastic and ataxic components with primary signs of slow rate, irregular speech timing, imprecise articulation, strained-strangled voiced and unnatural word stress expression.

In addition, 66 individuals comprised the controls cohort (41 females, 24 males) with a mean age of 45.5 (SD=11.2) years.

Each recording was represented by a feature vector of length 9, where 7 positions represent a diagnostically relevant acoustic feature of a speaker's recording, and 2 represent the age and gender of the speaker.

The overall workflow implemented for collecting, storing, initially analyzing to extracts 7 speech patter features and systematically building and evaluating the predictive models is depicted in Figure 1.

Acoustic Features Extraction and Validation

A resulting set of 7 acoustic features, listed in Table 1 and defined in the Appendix, were extracted from the delimited phonemes. These variables were used to distinguish between healthy and MS-afflicted individuals as it has been shown that changes in them are correlated with articulatory and cognitive impairments. Figure 2 includes some descriptive statistics for the 7-feature set extracted from the voice files.

Univariate and Multivariate Statistical Analyses of Variable significance

Next, we present the results of the univariate statistical significance testing using the Kolmogorov-Smirnoff (K-S) statistics, see Table 2. Only the 7 validated variables were included in the K-S testing, with five variables (listed in bold) were statistically significant or borderline significant (defined here as $p < \sim 0.1$).

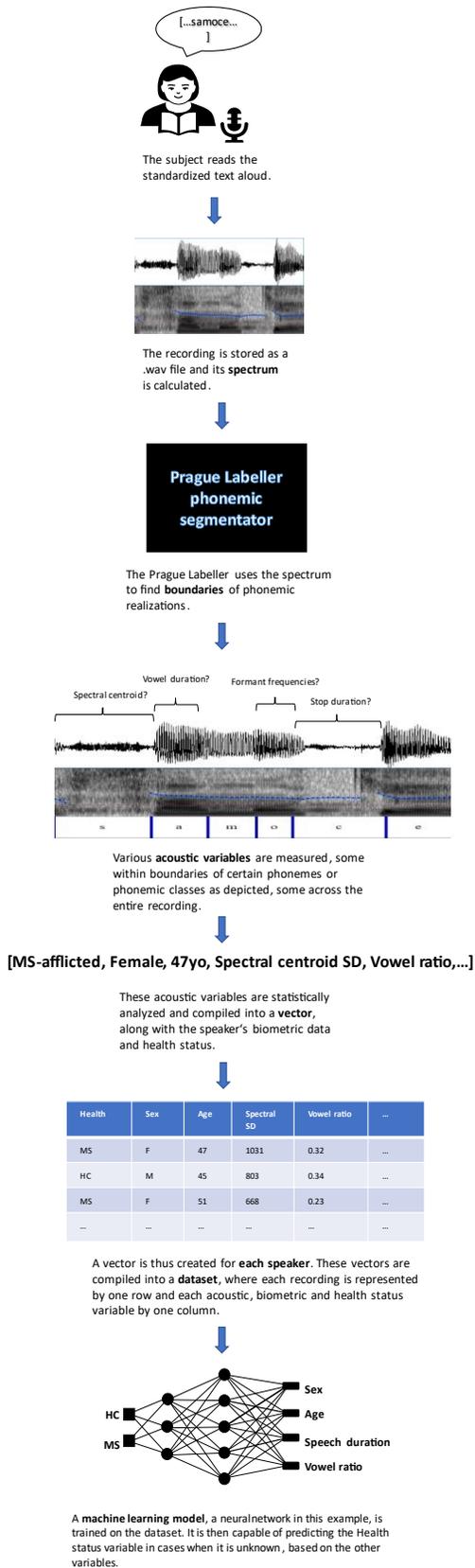

**Figure 1.**
A diagram of the feature extraction process workflow.

The Prague Labeller tool – shown as a black box in the diagram – is used to find points in the recordings when one individual speech sound ends and another begins. In the case of the Czech word *samotě* /samoce/, for instance, Prague Labeller uses the fact that /s/ represents a noisy sound as opposed to /a/, which represents a tonal sound, to find the boundary between these two at a certain point in time. This is called a *phonemic boundary* and comes attached with a label of its respective phoneme.

Some acoustic features are then measured within these boundaries, like the spectral centroid of /s/, which simultaneously roughly corresponds to the perceived "sharpness" of the sound and the configuration of the tongue while it is being pronounced. Similarly, acoustic features are measured across the *entire* recording, such as the CSI of $f_0$, which represents the total melody fluctuation across the recording.

These measured acoustic features, along with the given subject's sex and age, are compiled into vectors, which themselves are compiled into a dataset where each row represents one speaker with their corresponding acoustic features. One of the machine learning algorithms presented in Table 1, represented in the diagram using a neural network as an example, is then trained to predict the health status based on the biometric and acoustic data.

Thus, it is possible to train a model to predict the neurological health status of an individual using nothing but a .wav recording of them reading a predetermined text out loud.

Table 3 lists the results of the multi-variate statistically significant analyses after adjustment with all validated acoustic characteristics (variables) and gender and age, using a generalized linear regression model. Of note is that after adjustment, only two variables (CSI of vowel duration and Quantile difference of fundamental frequency) were borderline significant.

Classification Models

The results of the systematic model building, and evaluation are presented in Table 4. There were several models whose AUC achieved similar performance: eXGB, Gradient Boosting Machine, Random Forest and, to a degree, the Neural Network model. We chose to select the best-performing model based on a combination of both accuracy and the AUC measures (we used the rank of each model in each metric): it appears the best performing model was a Random Forest, which achieved an accuracy of 0.82 on a holdout validation dataset and an AUC of 0.76 as measured across 5 train/test cycles on the training data.

**Discussion**

In this study, we present a fully automated, quantitative assessment methodology to detect and objectively measure the footprint of the disease in the speech of MS patients. This opens the door for a scalable and unbiased diagnostic, disease progression and treatment response assessment of MS that can in principle aid the current standard clinical diagnostic assessment for MS (i.e., review of clinical history and examination, brain and spinal cord MRI, CSF analysis etc.; the McDonald Criteria (5)). Using only speech recordings and basic demographic factors, we have been successful in achieving an accuracy of 0.82 (AUC=0.76) using fully automated algorithmic methods. This points to the fact that MS carries a possible distinct vocal fingerprint which can in principle be utilized to diagnose or at least pre-diagnose the disease using methods orthogonal to the current standard-of care set of tests and procedures. Our study also corroborates some of the results of the landmark study published by Gerald et al. in 1987, (6) in which MS-related dysarthria was assessed auditorily only.

| Feature | p value |
|---|---|
| **Speech duration** | **0.008** |
| Vowel-to-recording ratio | 0.84 |
| **CSI of vowel duration** | **0.007** |
| **Quantile difference of $f_0$** | **0.007** |
| **Unvoiced stop mean duration** | **0.02** |
| **CSI of intensity** | **1.6 × 10$^{-7}$** |
| Spectral centroid of /s/, SD | 0.71 |

**Table 2**:
A table of the univariate statistics *p*-values calculated using the Kolmogorov-Smirnov test for the 7 variables extracted by the *Labeller* and validated against human expert annotations, see Methods and also Table 1. Parameters deemed significant or borderline statistically significant are in **bold**. "$f_0$" is a shorthand for fundamental frequency.

Because the results in our study have been measured objectively, they are much more rigorous; Gerald's study however covers other areas of affected linguistic capabilities, such as impaired grammar, which are much more non-equivocal to assess automatically. Therefore, finding a way to objectively assess such MS-related linguistic impairment beyond articulatory difficulties may present another set of indicators to help preliminary diagnosis of the disease.

Recently, a number of studies have attempted to use "real world" data (normally, clinical records data) to assess the risk of MS patient trajectories that transition from various established states in the MS disease progression, i.e. in the general disease course, (19) or for example the initial Relapsing-Remitting (RR) to the Secondary Progressive (SP) form of the disease. (20) It is of interest to see how the alternative set of variables extracted from the voice patterns as shown in this study could be added to these types of clinical predictive approaches to potentially enhance the accuracy of the resulting models.

| Feature | p value |
|---|---|
| Speech duration | 0.40 |
| Vowel-to-recording ratio | 0.82 |
| **CSI of vowel duration** | **0.10** |
| **Quantile difference of $f_0$** | **0.09** |
| Unvoiced stop mean duration | 0.64 |
| CSI of intensity | 0.53 |
| Spectral centroid of /s/, SD | 0.77 |
| Age | 0.42 |
| Gender | 0.61 |

**Table 3**:
A table of the *p* values of individual features (variables) after adjustment with the 7 validated acoustic features (see Table 1) and gender and age included, using a generalized linear model. The variables in **bold** are borderline significant. "$f_0$" is a shorthand for fundamental frequency.

This study brings to the forefront another interesting research subject, e.g. the use of acoustic feature characterization from voice as a proxy measure when trying to find anatomical correlates of the symptoms in brain magnetic resonance imaging (e. g. lesion-symptom mapping experiments). (21,22) For example, Rusz et al. (23) correlated specific articulatory difficulties with particular brain volume changes. Vavougios et al. (24) were able to produce a discriminant function equation able to reach an accuracy of 1.0 in discriminating MS patients from healthy controls based on electroglottographic data.

In 2020, Rozenstoks et al. (25) demonstrated that MS patients exhibit significant difficulty in performing certain tasks involving alternating syllables; Noffs et al. (26) built a unified acoustic speech score significantly correlated with cerebellar white matter volume and quality of life.

Analyzing Figure 2, MS speakers show greater variability than the healthy controls regarding speech duration. This is likely a consequence of pyramidal involvement due to widespread grey and white matter reductions in speakers with spastic dysarthria, leading to slower reading, which is consistent with Clark et al. (27) The generally lower pitch range of MS speakers, as shown by the fundamental frequency quantile difference, probably reflects low range of vibration frequencies resulting in monotonic speech. The overall greater mean durations of unvoiced stops of MS patients shows a tendency to hold full articulatory closures for an abnormally long time, indicating muscular spasticity of the tongue, which is consistent with the findings of Tykalová et al. (28) The two individuals exhibiting extreme high variability of intensity, as reflected in their f0 quantile difference values, pointing towards spasticity of the breathing muscles. Finally, articulation of the phoneme /s/ requires producing a shallow groove along the center of the tongue, which requires a high degree of coordination. The greater standard deviation of the phoneme's spectral centroid of MS speakers seems to reflect ataxia due to their compromised ability to consistently articulate the difficult phoneme.

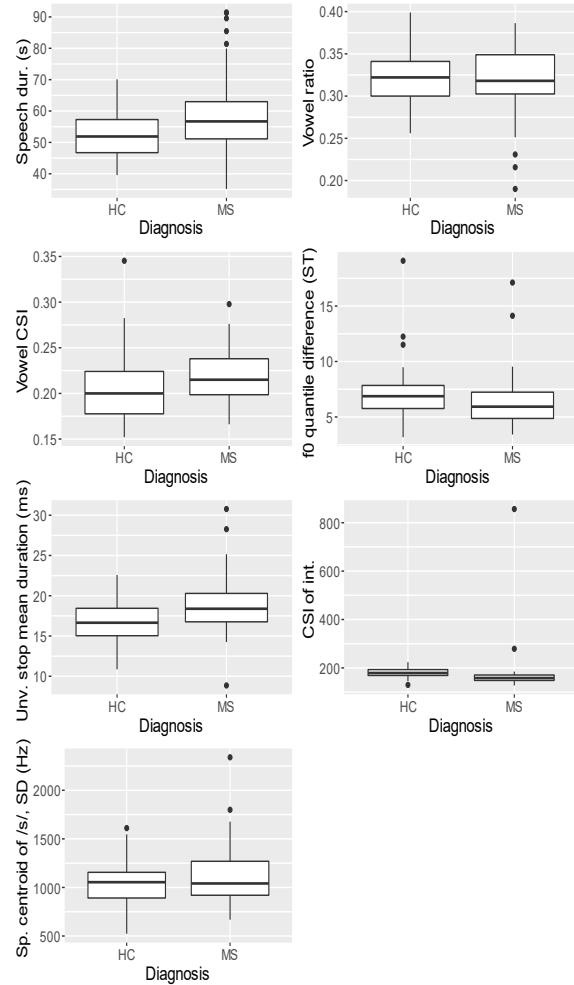

**Figure 2.**
Boxplots of the 12 acoustic measurements that were used as features. Note that recording and unvoiced durations are measured in milliseconds, the fundamental frequency measurements ("$f_0$") are in semitones (because fundamental frequency corresponds to articulatory phenomena logarithmically) and the formant measurements are listed in Hertz.

**Implications**

This study describes a fully automated, quantitative assessment methodology to detect and objectively measure the footprint of the disease in the speech of MS patients.

| Model | Accuracy | Sensitivity | Specificity | AUC |
|---|---|---|---|---|
| xGB | 0.70 | 0.73 | 0.67 | 0.79 |
| GBM | 0.77 | 0.77 | 0.77 | 0.75 |
| NN | 0.77 | 0.73 | 0.66 | 0.66 |
| **RF** | **0.82** | **0.90** | **0.75** | **0.76** |
| kNN | 0.58 | 0.67 | 0.56 | 0.66 |
| SVM | 0.46 | 0.45 | 0.47 | 0.66 |

**Table 4**:
A comparison of the Accuracy and AUC scores of each model as calculated on the validation dataset, and the mean area under the ROC curve as calculated across 5 resamples of cross-validation on the training set. The best performing model is in **bold**.

This opens the door for a scalable and unbiased diagnostic, disease progression and treatment response assessment of MS that can in principle be used in conjunction with and aid the current standard clinical diagnostic assessment for multiple sclerosis.

**Limitations**

Our study is not without limitations. First and foremost, it is possible that for a significant proportion of MS patients, there might be no dysarthria present at all. Theoretically, these zero-dysarthria patients should, however be relatively rare (at least as detected by acoustic analysis (19)), because speech motor control is governed by a high number of different areas of the CNS. (6,23) Related to this lack of diagnostic specificity to MS, this initial work has not established a precise map between MS disease progression stages and the speech patterns (variables) that are predictive of the disease status. This latter correspondence and its clinical validation need additional studies.

Additionally, the subjects have had in general MS for an average of 15 years, which is a rather long time.

Furthermore, the set of features and ML-models trained thereon used in this study has been limited to a specific language, Czech. Theoretically, there is nothing overly language-specific about this set of features and the workflow utilized, but additional studies and validation, including a clinical validation, are needed for English-speaking populations such as in the US, as well as other languages. We expect our results to extend to other languages, because it has been demonstrated that motor speech disorders can be acoustically measured cross-linguistically. (30)

Finally, our dataset is of rather limited size. Larger cohorts are desirable to validate and extend the findings reported here.

**Conclusions**

The use of machine learning and artificial intelligence in automated analyses of voice recording for aiding diagnosis, disease and treatment progression of multiple sclerosis holds promise. Further clinical validation of the specificity and a mapping to the MS disease progression phases is needed, as well as a validating utility for other languages.

Abbreviations: AUC (Area-Under-Curve), EDSS (Expanded Disability Status Scale), SP (Secondary Progressive), PP (Primary Progressive), RF (Random Forest) CSI (Cumulative Slope Index), SD (Standard Deviation), xGB (eXtreme Gradient Boosting), GLMnet (Generalized Linear Model), NN (Neural Net), RF (Random Forest) gbm (Generalized regression Boosting Model), KNN (k-Nearest Neighbors), SVM (Support Vector Machine).

**Declarations**

Ethics approval and consent to participate

The study was approved by the Ethics Committee of the General University Hospital in Prague, Czech Republic and have therefore been performed in accordance with the ethical standards laid down in the 1964 Declaration of Helsinki and its later amendments. All participants provided written, informed consent to the neurological examination and recording procedure.

Consent for Publication

Not Applicable

Availability of Data and material

Competing Interests

No financial or other, non-financial competing interest declared.

Funding

This study was supported by the Research Center for Informatics, grant nr. CZ.02.1.01/0.0/0.0/16_019/0000765, and by Czech Ministry of Education Grant PROGRESQ27/LF1.

Authors' Contributions

All authors have read and approved the manuscript

*Acknowledgements*

Not Applicable

# Appendix

**Speech duration**
The total duration of the recording in seconds, beginning with the first word and ending with the last, as delimited by either *Prague Labeller* or the human experts. Motivated by the assumption that trouble with muscle control would lead to articulatory difficulty, leading to longer recording times.

**Silence-to-speech ratio**
The total time spent by the speakers being silent divided by Speech duration.

Motivated by the assumption that speakers struggling with muscular stiffness would speak in short, labored bursts, and that speaker struggling with fatigue would pause frequently to rest.

**Vowel-to-speech ratio**
The total time spent by the speakers articulating vowels divided by Speech duration.

Motivated by the assumption that because vowels are easier to pronounce than consonants, speakers struggling with fatigue would spend more time articulating them.

**CSI of vowel duration**
Cumulative Slope Index is the absolute value of the sum of differences between each two consecutive elements of a vector of values, in this case the vector of durations of each vowel pronunciation across a given recording. It can be interpreted as a scalar describing the total rate of change of a variable over a series of steps in time. It is given by the formula

$$CSI(x) = \sum_{n}^{N-1} |x[n+1] - x[n]|$$

where *x* is the vector of values in question, *n* is the index of each element of that vector and *N* is the total length of the vector, according to Volín et al. (31)

Additionally, the entire sum can be divided by Speech duration, which is the case for all usages of *CSI* in this article. This is referred to as *normalized CSI*. Every *CSI* measurement used in this study has been normalized.

Motivated by the fact that abnormally high or low normalized *CSI* of vowel duration points to abnormal speech rhythm, suggesting neurological difficulties.

***CSI* of fundamental frequency**
*CSI* calculated from the *fundamental frequency* contour of the recording, or in other words, the vector of zeroth formant frequency values calculated at each time step, measured in semitones with a reference frequency of 100 Hz. It was measured using the autocorrelation method and octave jumps were not manually corrected. This may informally be interpreted as the *total rate of change in voice pitch* across the recording.

The time duration of these time steps were determined by *Praat*'s default setting, which applies to all *CSI* and formant measurements used in this paper.(32)

Motivated by the fact that poor control of the laryngeal muscles may contribute to an abnormal base frequency *CSI* value.

**Quantile difference of fundamental frequency**
The difference between the third quartile value of the base frequency vector and the first quartile value of the base frequency vector. Measured the same way as in the previous parameter. May be interpreted informally as the *voice pitch range* of the recording. Motivated by the fact that poor control of the laryngeal muscles may lead to an abnormally high or low quantile difference.

**Unvoiced stop mean duration**
The mean duration the speaker spent pronouncing the phonemes /p/, /t/, /k/, /c/, which are prototypically pronounced with a full closure of two articulatory organs. Motivated by the fact that individuals struggling with stiffness may hold this closure for a longer time.

**CSI of intensity**
CSI calculated from the *sound intensity* contour of the recording measured in decibels. May informally be interpreted as the total rate of change of speech loudness.

Motivated by the fact that fatigue may result in a flatter intensity contour due to breathing muscle weakness, leading to an unusually small CSI value; similarly, spasms of the breathing muscles may lead to a greater CSI value than normal.

**Standard deviation of the spectral centroid of /s/**
Standard deviation of the center of mass of the acoustic spectrum of all pronunciations of the /s/ phoneme across the recording. Can informally be understood from a perceptual viewpoint as the *overall variation in sound sharpness* when pronouncing this phoneme.

Motivated by the fact that the sharpness of /s/ is determined by the ability to produce and maintain a shallow groove in one's tongue whilst it is pressed against the roof of the mouth as a sufficiently strong flow of air from the lungs is maintained. Since this is a difficult task from a muscle coordination perspective, neurological difficulty may result in discrepancies.

**Standard deviations of $F_1$, $F_2$ and $F_3$**
Standard deviations of the frequencies of the first three spectral maxima (barring the base frequency) measured using the Burg method.

Motivated by the fact that the variability of these across time strongly correlates with the range of motion of the jaw, the range of motion of the tongue and overall muscle tenseness, which would all be strongly affected in an individual with MS.

**References**

1. Goldenberg MM. Multiple Sclerosis Review. Pharm Ther. 2012 Mar;37(3):175–84.

2. Dobson R, Giovannoni G. Multiple sclerosis - a review. Eur J Neurol. 2019 Jan;26(1):27–40.

3. Pavelek Z, Sobíšek L, Šarláková J, Potužník P, Peterka M, Štětkárová I, et al. Comparison of Therapies in MS Patients After the First Demyelinating Event in Real Clinical Practice in the Czech Republic: Data From the National Registry ReMuS. Front Neurol. 2021;11:1833.

4. McGinley MP, Goldschmidt CH, Rae-Grant AD. Diagnosis and Treatment of Multiple Sclerosis: A Review. JAMA. 2021 Feb 23;325(8):765–79.

5. Thompson AJ, Banwell BL, Barkhof F, Carroll WM, Coetzee T, Comi G, et al. Diagnosis of multiple sclerosis: 2017 revisions of the McDonald criteria. Lancet Neurol. 2018 Feb 1;17(2):162–73.

6. Gerald FJF, Murdoch BE, Chenery HJ. Multiple Sclerosis: Associated Speech and Language Disorders. Aust J Hum Commun Disord. 1987 Dec 1;15(2):15–35.

7. Rusz J. Detecting speech disorders in early Parkinson's disease by acoustic analysis. 2018;

8. Rusz J, Benová B, Růžičková H, Novotný M, Tykalová T, Hlavnicka J, et al. Characteristics of motor speech phenotypes in multiple sclerosis. Mult Scler Relat Disord. 2018 Jan;19:62–9.

9. Hartelius L, Runmarker B, Andersen O. Prevalence and Characteristics of Dysarthria in a Multiple-Sclerosis Incidence Cohort: Relation to Neurological Data. Folia Phoniatr Logop. 2000;52(4):160–77.

10. Noffs G, Boonstra FMC, Perera T, Butzkueven H, Kolbe SC, Maldonado F, et al. Speech metrics, general disability, brain imaging and quality of life in multiple sclerosis. Eur J Neurol. 2021 Jan;28(1):259–68.

11. Polman CH, Reingold SC, Banwell B, Clanet M, Cohen JA, Filippi M, et al. Diagnostic criteria for multiple sclerosis: 2010 Revisions to the McDonald criteria. Ann Neurol. 2011;69(2):292–302.

12. Kurtzke JF. Rating neurologic impairment in multiple sclerosis: An expanded disability status scale (EDSS). Neurology. 1983 Nov 1;33(11):1444–1444.

13. Rusz J, Tykalova T, Ramig LO, Tripoliti E. Guidelines for Speech Recording and Acoustic Analyses in Dysarthrias of Movement Disorders. Mov Disord. 2021;36(4):803–14.

14. Čapek K. Měl jsem psa a kočku [Internet]. Městská knihovna v Praze; 1939 [cited 2020 May 16]. Available from: https://search.mlp.cz/cz/titul/mel-jsem-psa-a-kocku/3347549/

15. Pollák P, Volín J, Skarnitzl R. HMM-based phonetic segmentation in Praat environment. In: Proceedings of the VII th International Conference "Speech and Computer–SPECOM. 2007. p. 537–41.

16. Machač P, Skarnitzl R. Fonetická segmentace hlásek. [Internet]. Epocha; 2010. Available from: https://search.ebscohost.com/login.aspx?authtype=shib&custid=s1240919&profile=eds

17. Kuhn M. topepo/caret [Internet]. 2020 [cited 2020 May 10]. Available from: https://github.com/topepo/caret

18. Kuhn M. Building Predictive Models in R Using the caret Package. J Stat Softw. 2008 Nov 10;28(1):1–26.

19. Zhao Y, Wang T, Bove R, Cree B, Henry R, Lokhande H, et al. Ensemble learning predicts multiple sclerosis disease course in the SUMMIT study. Npj Digit Med. 2020 Oct 16;3(1):1–8.

20. Seccia R, Romano S, Salvetti M, Crisanti A, Palagi L, Grassi F. Machine Learning Use for Prognostic Purposes in Multiple Sclerosis. Life. 2021 Feb;11(2):122.

21. Bates E, Wilson SM, Saygin AP, Dick F, Sereno MI, Knight RT, et al. Voxel-based lesion-symptom mapping. Nat Neurosci. 2003 May;6(5):448–50.

22. Wilson SM. Lesion-symptom mapping in the study of spoken language understanding. Lang Cogn Neurosci. 2017;32(7):891–9.

23. Rusz J, Vaněčková M, Benová B, Tykalová T, Novotný M, Ruzickova H, et al. Brain volumetric correlates of dysarthria in multiple sclerosis. Brain Lang. 2019;194:58–64.

24. Vavougios GD, Doskas T, Konstantopoulos K. An electroglottographical analysis-based discriminant function model differentiating multiple sclerosis patients from healthy controls. Neurol Sci. 2018 May 1;39(5):847–50.

25. Rozenstoks K, Novotný M, Horáková D, Rusz J. Automated Assessment of Oral Diadochokinesis in Multiple Sclerosis Using a Neural Network Approach: Effect of Different Syllable Repetition Paradigms. IEEE Trans Neural Syst Rehabil Eng Publ IEEE Eng Med Biol Soc. 2020;28(1):32–41.

26. Noffs G, Boonstra FMC, Perera T, Kolbe SC, Stankovich J, Butzkueven H, et al. Acoustic Speech Analytics Are Predictive of Cerebellar Dysfunction in Multiple Sclerosis. Cerebellum Lond Engl. 2020 Oct;19(5):691–700.

27. Clark HM, Duffy JR, Whitwell JL, Ahlskog JE, Sorenson EJ, Josephs KA. Clinical and imaging characterization of progressive spastic dysarthria. Eur J Neurol. 2014;21(3):368–76.

28. Tykalova T, Rusz J, Klempir J, Cmejla R, Ruzicka E. Distinct patterns of imprecise consonant articulation among Parkinson's disease, progressive supranuclear palsy and multiple system atrophy. Brain Lang. 2017 Feb 1;165:1–9.

29. Klevan G, Jacobsen CO, Aarseth JH, Myhr K-M, Nyland H, Glad S, et al. Health related quality of life in patients recently diagnosed with multiple sclerosis. *Acta Neurol Scand*. 2014 Jan 1;129(1):21–6.

30. Rusz J, Hlavnička J, Novotný M, Tykalová T, Pelletier A, Montplaisir J, et al. Speech Biomarkers in Rapid Eye Movement Sleep Behavior Disorder and Parkinson Disease. Ann Neurol. 2021;90(1):62–75.

31. Volín J, Tykalová T, Boril T. Stability of Prosodic Characteristics Across Age and Gender Groups. In 2017. p. 3902–6.

32. Time step settings... [Internet]. [cited 2021 Feb 24]. Available from: https://www.fon.hum.uva.nl/praat/manual/Time_step_settings___.html